\documentclass[prd,superscriptaddress]{revtex4}
\usepackage{graphicx}
\usepackage{epsf}
\usepackage{epsfig}
\usepackage{amssymb}
\usepackage{amsmath}

\def\1{{\'{\i}}}
\newcommand{\ds}{\displaystyle}
\newcommand{\df}{\displaystyle\frac}
\newcommand{\be}{\begin{equation}}
\newcommand{\ee}{\end{equation}}
\newcommand{\bea}{\begin{eqnarray*}}
\newcommand{\eea}{\end{eqnarray*}}

\begin{document}
\title{Dynamics of Interacting Quintessence Models: Observational Constraints}
\author{Germ\'{an} Olivares\footnote{E-mail address: german.olivares@uab.es}}
\affiliation{Departamento de F\'{\i}sica, Universidad Aut\'{o}noma de Barcelona,
Barcelona, Spain}
\author{Fernando Atrio-Barandela\footnote{E-mail address: atrio@usal.es}}
\affiliation{Departamento de F\'{\i}sica Fundamental,
Universidad de Salamanca, Spain}
\author{Diego Pav\'{o}n\footnote{E-mail address: diego.pavon@uab.es}}
\affiliation{Departamento de F\'{\i}sica, Universidad Aut\'{o}noma de Barcelona,
Barcelona, Spain}

\begin{abstract}
Interacting quintessence models have been proposed to explain or,
at least, alleviate the coincidence problem of late cosmic
acceleration. In this paper we are concerned with two aspects of
these kind of models: (i) the dynamical evolution of the model of
Chimento {\em et al.} [L.P. Chimento, A.S. Jakubi, D. Pav\'{o}n,
and W. Zimdahl, Phys. Rev. D \textbf{67}, 083513 (2003).], i.e.,
whether its cosmological evolution gives rise to a right sequence
of radiation, dark matter and dark energy dominated eras, and (ii)
whether the dark matter dark energy ratio asymptotically evolves
towards a non-zero constant. After showing that the model
correctly reproduces these eras, we correlate three data sets that
constrain the interaction at three redshift epochs: $z\le 10^{4}$,
$z=10^{3}$, and $z=1$. We discuss the model selection and argue
that even if the model under consideration fulfills both
requirements, it is heavily constrained by observation. The
prospects that the coincidence problem can be explained by the
coupling of dark matter to dark energy are not clearly favored by
the data.
\end{abstract}
\maketitle

\section{Introduction}
Recent measurements of luminosity distances using supernovae type
Ia (SNIa) \cite{riess}, of the cosmic microwave background (CMB)
temperature anisotropies with the Wilkinson Microwave Anisotropy
(WMAP) satellite \cite{wmap3}, large scale structure \cite{lss},
the integrated Sachs--Wolfe effect \cite{isw}, and weak lensing
\cite{weakl}, strongly suggest that the Universe is currently
undergoing a phase of accelerated expansion -see \cite{reviews}
for recent reviews. Within general relativity the obvious
candidate to explain the present acceleration is the cosmological
constant (or vacuum energy), and in fact the na\"{i}ve model built
on it, $\Lambda$CDM, seems to pass reasonably well all
cosmological tests. However, it suffers from two serious drawbacks
from the theoretical side: the unnatural low value of the
corresponding energy density, 123 magnitude orders larger than
observed, and the so-called ``coincidence problem", namely, ``Why
are the densities of matter and vacuum of the same order precisely
today?", that requires the vacuum energy density to be 96 orders
of magnitude smaller than the matter density at the Planck scale.
(It is fair, however, to mention the existence of proposals in
which a vacuum energy density of about the right order of
magnitude stems from the Casimir effect at cosmic scales -see
\cite{emili} and references therein). This is why much attention
has been devoted to models featuring an evolving and nearly
un-clustered form of energy, usually dubbed ``dark energy ",
possessing a strong negative pressure high enough to drive late
acceleration -see \cite{edmund} for an ample review of models.

For simplicity, most cosmological models assume that matter and
dark energy interact gravitationally only. In the absence of an
underlying symmetry that would suppress a matter-dark energy
coupling (or interaction) there is no a priori reason to dismiss
it. Further, the coupling is not only likely but inevitable
\cite{jerome} and its introduction is not more arbitrary than
assuming it to vanish. On the other hand, it may help explain the
coincidence problem. Ultimately, this question will have to be
resolved observationally. Among other things, the interaction can
push the beginning of the era of accelerated expansion to higher
redshifts and it may erroneously suggest (if the interaction is
ignored when interpreting  the data) an equation of state for the
dark energy of phantom type -see \cite{amendola-gasperini} and
references therein. Another question, is the form of the coupling.
There is no clear consensus on this point and different versions,
that arise from a variety of motivations, coexist in the
literature.

Cosmological models where dark matter (DM) and dark energy (DE) do
not evolve separately but interact with each other were first
introduced to justify the small current value of the cosmological
constant \cite{wetterich} and nowadays there is a growing body of
literature on the subject -see, e.g. \cite{list} and references
therein. Recently, various proposals at the fundamental level,
including field Lagrangians, have been advanced to account for the
coupling \cite{piazza}. Lagrangians of scalar field coupled to
matter generically do not generate scaling solutions with a dark
matter dominated period lasting long enough as required by cosmic
structure formation \cite{amendola-challenges}.

In this paper we compare the interacting quintessence model (IQM)
of Chimento {\em et al.} \cite{chimento} (see also its forerunner
\cite{zimdahl}) with observational data (supernovae, cosmic
microwave background (CMB), and matter power spectrum) to set
limits on the strength of the interaction DM/DE. The model was
built to simultaneously account for the late phase of acceleration
in the framework of Einstein relativity and significantly
alleviates the coincidence problem. It evades  the limits set in
\cite{amendola-challenges} and  is compatible with a right
succession of cosmic eras -radiation, dark matter, and dark energy
dominated expansions. In a recent paper, Guo {\em et al.} also set
constraints on interacting quintessence models \cite{guo}.
However, the interactions studied by these authors differ from the
one in \cite{chimento}, and while they use the cosmic background
shift parameter and baryon acoustic oscillations alongside
supernovae data to constrain the interaction, they do not consider
the matter power spectrum, whereby our analysis may be viewed as
complementary to theirs.

The outline of this paper is as follows. Next section studies the
critical points of the autonomous system of equations associated
to the IQM \cite{chimento}. Section III considers the restrictions
set  by Amendola {\em et al.} on the model to conclude that  the
latter evades these restrictions and, in particular, that an early
epoch of baryon dominance is possible only if the strength of
interaction is unnaturally large (beyond the limits set by the CMB
data).  Section IV focus on the observational bounds coming from
the CMB, matter power spectrum and recent supernovae type Ia data.
Finally, section V summarizes our results.

\section{Dynamics of the Interacting Quintessence Model}
If the quintessence DE  decays into DM, both energy densities
evolve differently than in non-interacting quintessence
cosmologies and, therefore, the interaction can be tested by its
effects on the dynamical evolution of the Universe. Due to the
interaction, the fraction of cold dark matter (CDM) at any time in
the past is smaller than in non-interacting models with the same
cosmological parameters today. Since the dark matter energy
density grows more slowly the beginning of the period of
accelerated expansion and the temporal evolution of the
gravitational potential differ from non-interacting models.
Observables such as the angular and luminosity distances depend on
the time evolution of the energy density. But the effect does not
only occurs at zeroth order; the evolution of first order matter
density perturbations is also affected and so is the pattern of
anisotropies of the CMB. This section describes the dynamical
evolution of dark matter and dark energy densities at zeroth order
in the interacting quintessence model of Ref. \cite{chimento} to
point out the main differences with respect to models with no
interaction.

The phenomenological model of Chimento {\em et al.}
\cite{chimento} (see also \cite{zimdahl}) assumes that the dark
energy decays into cold dark matter thereby both energy densities
do not longer evolve separately. In \cite{olivares1} the following
ansatz was adopted
\\
\begin{equation}\label{cont}
\dot\rho_c +3H\rho_c = 3 H \, c^2(\rho_x+\rho_c)\, , \qquad \quad
\dot\rho_x +3H(1+w_x)\rho_x = -3 H \, c^2(\rho_x+\rho_c), \
\end{equation}
\\
where $\rho$ denotes the energy densities (subscript $c$ for cold
dark matter and subscript $x$ for dark energy), $w_x$ is the
equation of state parameter of dark energy, $H \equiv a^{-1}da/dt$
the Hubble function and $a$ the scale factor of the flat
Friedmann-Robertson-Walker metric. The model assumes that the dark
energy decays just into cold dark matter and not into any other
component such as neutrinos, baryons, or photons. Decay into
neutrinos have been recently studied in the literature
\cite{brookfield}. The coupling with baryons is constrained by
measurements of local gravity \cite{peebles-rmph,hagiwara}. The
above ansatz is essentially phenomenological; it is the simplest
interaction model that leads to a  constant ratio at early and
late times which certainly alleviates the coincidence problem.
While at present a rigorous derivation  from first principles does
not exist, the ansatz may be roughly justified as follows: the
right hand side of both equations in (\ref{cont}) must be
functions, say $Q$ and $-Q$, of the energy densities multiplied by
a quantity with units of inverse of time. For the latter the
obvious choice is the Hubble factor $H$, so we have that $Q = Q(
H\rho_{x}\, , H\rho_{c})$. By power law expanding $Q$ and
retaining just the first term we get $Q \simeq \lambda_{x}\, H
\rho_{x} + \lambda_{c}\, H \rho_{c}$. To facilitate comparison of
the model with observation it is suitable to eliminate one the two
$\lambda$ parameters. Thus we  set $\lambda_{x} = \lambda_{c} = 3
c^{2}$ and arrive to Eq. (\ref{cont}). The simpler choice
$\lambda_{c}=0$ would not yield a constant dark matter to dark
energy ratio at late times. The dimensionless term $3 c^{2}$
measures to what extent the decay rate differs from the expansion
rate of the Universe. Thus, the model is characterized by this
single parameter, $c^{2}$, which also gauges the intensity of the
interaction. The lower $c^2$, the closer the evolution of the
Universe to a non-interacting model is.

Equations in (\ref{cont}), the continuity equations for photons
and baryons  and Friedmann  equation  $H^{2} =
(\kappa^{2}/3)(\rho_{rad} + \rho_{b}+ \rho_{c}+\rho_{x})$,
constitute a closed system.  In Fig. \ref{fig0} we plot the
evolution of the different matter density components: cold dark
matter (thick solid line), dark energy (thin solid), photons (thin
dot-dashed) and baryons (thick dot-dashed) for two values of the
interaction parameter: left and right panels correspond to
$c^2=10^{-3}$ and $c^2=0.1$, respectively. In the figure we can
see that Eqs.~(\ref{cont}) have two asymptotic solutions with $r=$
constant: at early times and in the immediate future. It is easy
to prove that, in these two cases, $r^2+(w_xc^{-2}+2)r+1=0$ and
the constant ratio is $r_-=r(z\rightarrow\infty)\sim c^{-2}$ and
$r_+=r(z\rightarrow 0)\sim c^2$. These results represent a clear
conceptual advantage with respect to the $\Lambda$CDM model. If
the present acceleration of the Universe is generated by a
cosmological constant, then the initial values of dark matter and
cosmological constant have to be tuned by 96 orders of magnitude
at Planck time, i.e., $r(t_{Planck})=\Omega_m/\Omega_\Lambda\sim
{\cal O}(10^{96})$. In the model described above this ratio is
fixed by the interaction rate, $c^2$. The initial condition
problem is significantly alleviated.  For example, if $c^2\sim
10^{-4}$ the initial dark matter to dark energy ratio will be
$r(t_{Planck})\sim{\cal O}(10^4)$.

To study the evolution of the corresponding autonomous system
formed by the continuity equations of all energy components and
Friedmann's equation we introduce the following set of variables,
\\
\begin{equation}
x= \df{\kappa}{H}\sqrt{\df{\rho_{x}}{3}}\, ,\;\; y= \df{\kappa}{H}
\sqrt{\df{\rho_{c}}{3}}\, , \;\; z =
\df{\kappa}{H}\sqrt{\df{\rho_{b}}{3}} \, , \;\; u=
\df{\kappa}{H}\sqrt{\df{\rho_{rad}}{3}} \, ,\label{newvar}
\end{equation}
\\
with $\kappa=\sqrt{8\pi G}$. Then, the dynamical equations can be recast as
\\
\begin{equation}\label{contnew}
\begin{array}{rcl}
x^{\prime}&=&\df{1}{2}\left[3w_{x}\left(x^2-1\right)+u^2-
        3c^2\ds\left(1+\df{y^2}{x^2}\right)\right]x \, ,\vspace*{.2cm}\\
y^{\prime}&=&\df{1}{2}\left[3w_{x}x^2+u^2+3c^2\ds\left(1+\df{x^2}{y^2}\right)\right]y
    \, ,\vspace*{.2cm}\\
z^{\prime}&=&\df{1}{2}\left[3w_{x}x^2+u^2\right]z \, ,\vspace*{.2cm}\\
u^{\prime}&=&\df{1}{2}\left[3w_{x}x^2+u^2-1\right]u\, ,\vspace*{.2cm}\\
1 &=& x^2+y^2+z^2+u^2 \, , \vspace*{.2cm}\\
\end{array}
\end{equation}
\\
where a prime denotes derivative with respect to $\ln a$. As said
above, the motivation for interacting quintessence models is to
solve or, at least, ameliorate the coincidence problem. A solution
will be achieved if the system (\ref{contnew}) presents scaling
solutions \cite{CopLid}. As already discussed, scaling solutions
are characterized by a constant dark matter to dark energy ratio.
Even more important are those scaling solutions that are also an
attractor. In this way, the coincidence problem gets substantially
alleviated because, regardless of the initial conditions, the
system evolves toward a final state where the ratio of dark matter
to dark energy stays constant.

\begin{table}\centering
\begin{tabular}{|p{3.5cm}|p{3.5cm}|p{3.5cm}|p{3.5cm}|}
\hline
$(x_c^2,y_c^2,z_c^2,u_c^2)$ & Existence condition    &Stability character  & Acceleration \\
\hline
$(x_{-}^2,y_{-}^2,0,0)$ & $0<c^2<|w_{x}|/4 $ &attractor &$\ddot a>0$ if $w_{x}<-1/3$\\
\hline
$(x_{+}^2,y_{+}^2,0,0)$ & $0<c^2<|w_{x}|/4 $ &saddle point &$\ddot a<0 \; \forall \,c^2,w_{x}$\\
\hline
$(0,0,1,0)$  & $\forall\, c^2,w_{x}$ &unstable  & $\ddot a<0$\\
\hline
$(0,0,0,1)$  & $\forall\, c^2,w_{x}$ &unstable  & $\ddot a<0$ \\
\hline
\end{tabular}
\normalsize
\medskip
\caption{Location of the critical points of the autonomous system
of Eqs. (\ref{contnew}), their stability and  dynamical behavior
of the Universe at those points.} \label{table:criticalpoints}
\end{table}

In Table \ref{table:criticalpoints} we give the location of the
four critical points of the autonomous system of Eqs.
($\ref{contnew}$), their stability character and whether
they give rise to accelerated or decelerated expansions. The first
critical point, $(x_{-}^2,y_{-}^2,0,0)$, corresponds to the dark
energy dominated epoch with
\\
\begin{equation}
x_{-}^2 = \df{1}{2w_x}\left(w_x-\sqrt{w_x^2+4c^2w_x}\right) \, ,
\qquad y_{-}^2 =
1-\df{1}{2w_x}\left(w_x-\sqrt{w_x^2+4c^2w_x}\right)\, ,
\end{equation}
\\
and it requires $c^2< |w_{x}|/4$. As shown in \cite{olivares1},
this coincides with the condition the interaction must fulfill to
give a physically acceptable evolution for the Universe.

The second critical point,
\\
\begin{equation}
x_{+}^2 = \df{1}{2w_x}\left(w_x+\sqrt{w_x+4c^2w_x}\right) \, ,
\qquad y_{+}^2 =
1-\df{1}{2w_x}\left(w_x+\sqrt{w_x^2+4c^2w_x}\right)\, ,
\end{equation}
\\
corresponds to the CDM dominated era and it is a saddle point.

The third critical point, $(0,0,1,0)$, is unstable and physically
unrealistic. It corresponds to a universe with just baryons (it
contains neither radiation, nor matter, nor dark energy). Finally,
the fourth point, $(0,0,0,1)$, corresponds to the radiation
dominated era and it is also unstable. To summarize, as soon as
the other energy densities grow more important than radiation, the
system moves away from this point, reaches a CDM dominated era,
which corresponds to a saddle point, and eventually moves to a
final attractor scaling solution in the dark energy dominated era.
In a natural way, the dynamical evolution of the Universe is to
asymptotically approach this attractor,  a never-ending phase of
accelerated expansion, in which the energy ratio $r \equiv
\rho_{c}/\rho_{x}$ remains constant.

\section{IQM and the coincidence problem}
In \cite{amendola-challenges} Amendola {\em et al.} discussed the
conditions that interacting quintessence models must verify in
order to solve the coincidence problem  and have a correct
sequence of cosmological eras -i.e., radiation, matter, and
accelerated scaling solution.  Many of the Lagrangians proposed in
the literature do not correspond to acceptable scaling solutions.
Those models either enter a phase of accelerated expansion right
after the radiation epoch or present baryon dominated periods that
would affect the growth of structure in the Universe. This brings
about a great stress on the ability of a large class of models  to
explain or even ameliorate the coincidence problem.

\begin{figure}[t]
\centering\epsfig{file=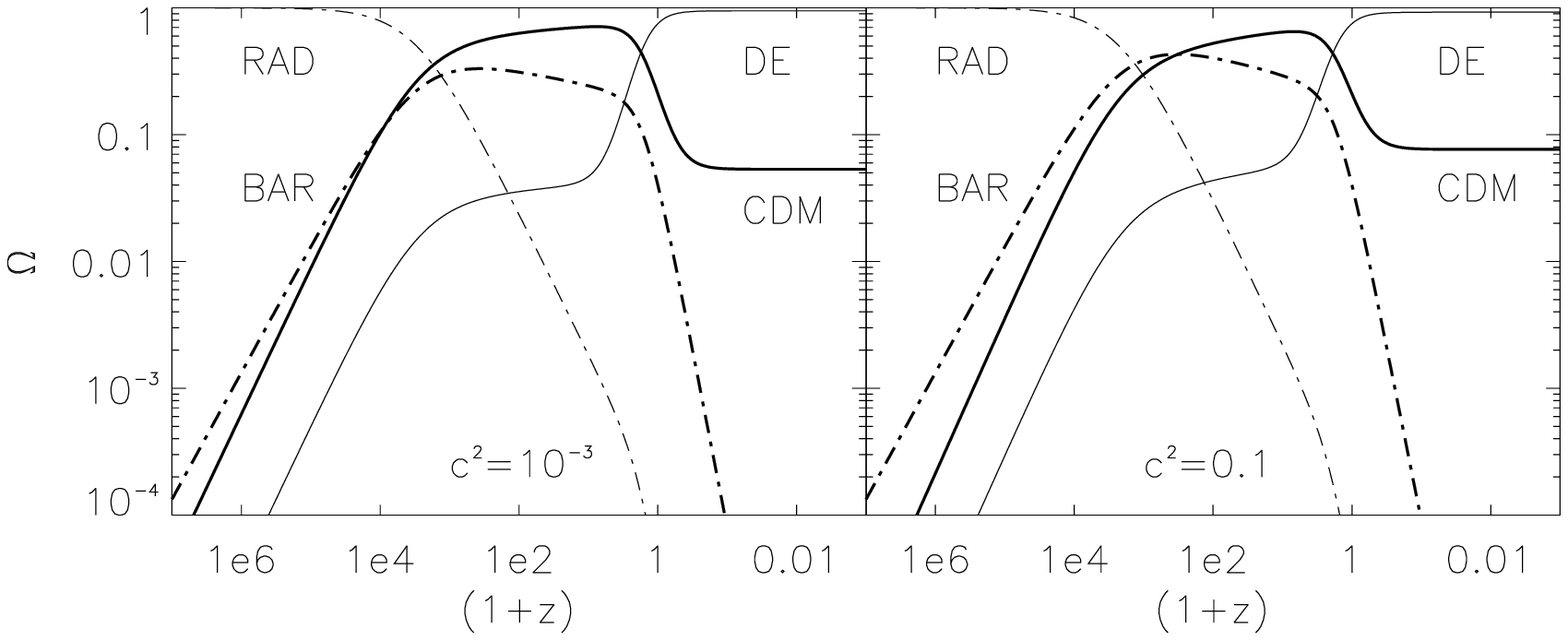,width=17cm} \vspace*{-6.0cm}
\caption{Evolution of the different energy densities with
redshift. Left panel: $c^2=10^{-3}$, right panel: $c^2=0.1$. Thick
and thin solid, and thick, and thin dot-dashed lines correspond to
dark matter, dark energy, baryons and photons, respectively. }
\label{fig0}
\end{figure}

The phenomenological model of Chimento {\em et al.}
\cite{chimento} is not affected by the results of Ref.
\cite{amendola-challenges} since the coupling between dark matter
and dark energy is not of the type studied by Amendola {\em et
al.}, namely, $Q \propto \rho_{m}\, d\phi/d(\ln a)$, where
$\rho_m$ denotes the dark matter energy density and $\phi$ the
scalar field. Equivalently, the Lagrangian associated to the
effective potential given in Eq. (5) of Ref. \cite{olivares1} is
not of the type described in \cite{amendola-challenges}. As can be
seen in Fig. \ref{fig0} the model contains two scaling regime
solutions, one in the radiation- and matter-dominated eras,
another in the period of accelerated expansion -at late times. It
reproduces the right succession of radiation, matter and
accelerated epochs, each lasting long enough to give rise to the
observed cosmic structure. It only remains to show under which
circumstances there is not a baryon dominated period  (as in left
panel of Fig. \ref{fig0}) so that the formation of galaxies and
large scale structure go unaffected.

\subsection{Baryon dominated era}
Some interacting quintessence models present the severe drawback
that the Universe undergoes a baryon-dominated period  prior to
the DM era. This is rather problematic since CMB temperature
anisotropies and the evolution of matter density perturbations
would result largely affected. Unlike the models of Ref.
\cite{tocchini}, the Chimento {\em et al.} model \cite{chimento}
does not have a stable baryon dominated era. To see this let us
estimate when and for how long could the model we are considering
undergoes a baryon dominated epoch. The maximum baryon energy
density can be obtained by setting $z^{\prime}= 0$ in  Eqs.
(\ref{contnew}). Noting that $u^2=-3w_x\, x^2$, we have from the
last equation in (\ref{contnew}) that
\\
\begin{equation}\label{bar_max}
z^2_{max}=1-y^2-(1-3w_x)x^2\, .
\end{equation}
\\
For $r=$ constant, $r^2+(w_{x} \, c^{-2}+2)r+1 = 0$. To affect the
growth of structure and the  CMB temperature anisotropies, baryon
domination should occur, if at all, well before the period of
accelerated expansion, i.e., when   $1\ll r\simeq r_{+}$ with
$r_{+} = (1/2)[-(w_{x} \, c^{-2} \, + 2) + \sqrt{(w_{x} \, c^{-2}
\, + 2)^{2}\, -4}] \simeq |w_{x}| c^{-2} $. Since $y^2 = r \, x^2$
Eq. (\ref{bar_max}) becomes
\\
\begin{equation}
z^2_{max}\simeq 1-(1+3c^2)y^2\, .
\end{equation}
\\
Therefore, for an early baryon dominated era to have existed we
must have $z^2_{max}\simeq 1-(1+3c^2)y^2 > x^2+y^2$. Thus, bearing
in mind that $x^{2} \ll y^{2}$ the condition for the said epoch to
exist is
\\
\begin{equation}
y^2 \le \df{1}{2+3c^2}\, .
\end{equation}
\\
A baryon dominated epoch may have occurred for $c^{2}\sim {\cal
O}(10^{-1})$ or larger. However, this does not correspond to any
stable critical point of the autonomous system and -as we shall
see below- such high values of $c^{2}$ are ruled out by the CMB
data. Therefore, for $c^{2}$ values of about $10^{-2}$ or lower
the model of Ref. \cite{chimento} is free of a long period of
baryon dominance.

To conclude, the succession of expansion eras in the model we are
considering can be summarized as follows: after the initial
radiation dominated era, the Universe might have gone through a
very short baryon dominated period (if at all) followed by a DM
dominated epoch (corresponding to an unstable critical point), to
finally asymptotically approach a regime with a constant dark
matter to dark energy ratio, at late times, in the  accelerated
expansion era.

\section{Comparison with observational data}
The IQM of Ref. \cite{chimento} evades the restrictions imposed by
\cite{amendola-challenges} and provides a viable model to explain
the coincidence problem by means of cosmological scaling
solutions. The actual existence of an interaction must be
elucidated by contrasting the model with observations. In Refs.
\cite{olivares1} and \cite{olivares2} we studied the effect of the
interaction on the temperature anisotropies of the CMB, on matter
power spectrum and on the luminosity distance. This opens up the
possibility of using observational data to constrain the strength
of the interaction -i.e., the $c^{2}$ parameter-, at different
redshifts. At present, data are available on the power spectrum of
matter density perturbations extracted from galaxy catalogs
\cite{2df,sdss,sdss2}, on temperature anisotropies ranging from
the largest scales down to $\sim 20$ arcmin \cite{wmap3} and on
luminosity distances up to redshift $z\sim 1$ measured using SNIa
as standard candles \cite{riess}. Each data set probes the DE/DM
coupling at different periods: CMB anisotropies at $z\simeq 10^3$,
the slope of the matter power spectrum at $z\le 10^4$ and the
luminosity distance test at $z\sim 1$ and below. We carried out
independent statistical analysis for each data set to see what
constraints did the data impose on the interaction at different
epochs.

\begin{figure}[t]
\centering\epsfig{file=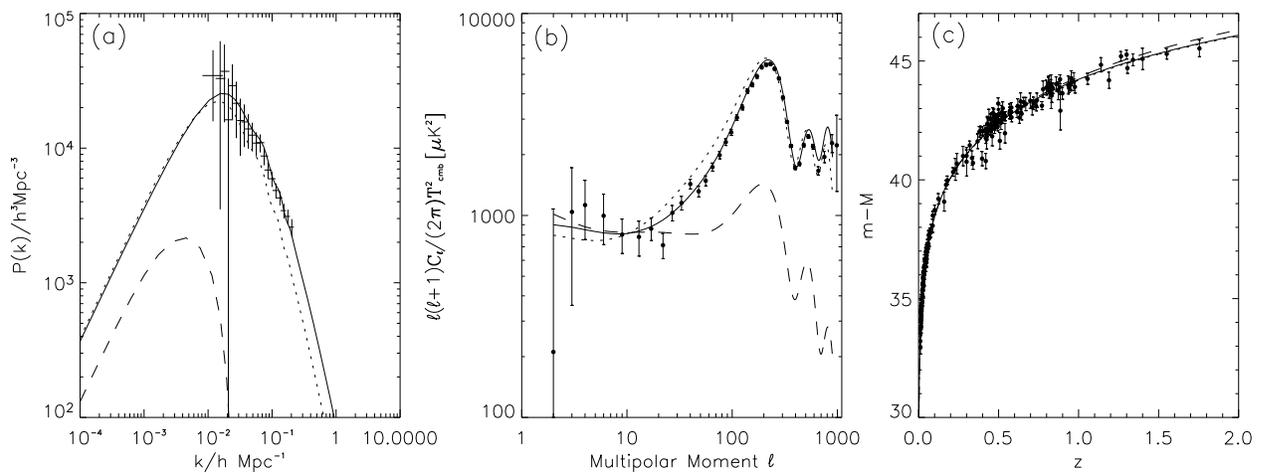,width=17cm} \vspace*{-2cm}
\caption{Matter, radiation power spectrum, and luminosity distance
for a cosmological model with $\Omega_{x}=0.74$ and $w_x=-0.9$ and
three different interaction parameters $c^2= 0, 10^{-2}, 0.1$,
corresponding to solid, dotted, and long dashed lines,
respectively. The cosmological parameters are those of the
fiducial WMAP 3yr data. } \label{fig:cmb3rd}
\end{figure}

The effect of the interaction on the various observables can be
determined numerically only. Physically, the interaction renders
the potential wells shallower in the matter dominated regime and
matter density perturbations grow slower \cite{olivares2}; the
gravitational pull on baryons prior to recombination weakens and
the interaction alters the relative height of the acoustic peaks
of the CMB spectrum and, finally, the interaction also changes the
time evolution of the different energy densities, i.e., it affects
the luminosity distance of standard candles. Figure
\ref{fig:cmb3rd} shows these effects. In Figure
\ref{fig:cmb3rd}(a)-\ref{fig:cmb3rd}(c) we plot the matter,
radiation power spectra, and the luminosity distance for three
cosmological models with the parameters of WMAP 3yr data
concordance model and interaction parameter $c^2=0, 10^{-2}, 0.1$
corresponding to solid, dotted and long dashed lines,
respectively. We also plot the data (see below) used in the
statistical analysis. In Fig. \ref{fig:cmb3rd}(a), the effect of
the interaction is to shift the matter radiation equality and
strongly damp the slope of the matter power spectrum. As a result,
a large interaction erases the matter power spectrum at small
scales. In the figure, power spectra are normalized to Cosmic
Microwave Background/Differential Microwave Radiometers
(COBE/DMR). In Fig. \ref{fig:cmb3rd}(b) the interaction changes
the relative height of the acoustic peaks. At low multipoles the
amplitudes also differ, since the slower evolution of the
potential at low redshift gives rise to an Integrated Sachs-Wolfe
component of different amplitude. This effect is discussed in
\cite{olivares4}. In Fig. \ref{fig:cmb3rd}(c) we see that the
effect of the interaction is not very significant, and only when
$c^{2}$ is as high as $\sim 0.1$ there is a tiny difference with
non-interacting models that can be perceived at $z \sim 2$.

To constrain the model with observations, for every given set of
cosmological parameters we generate the power spectrum of matter
density perturbations and temperature anisotropies of the CMB
using a modified version of the CMBFAST code \cite{cmbfastweb}.
Our cosmological models are described by the following parameters:
dark matter, baryon and dark energy densities ($\Omega_{c}$,
$\Omega_b$, $\Omega_x = 1- \Omega _{c}-\Omega_{b}$), dark energy
equation of the state parameter ($w_x$), Hubble constant ($H_0$),
amplitude and slope of the matter power spectrum at large scales
($A_S$, $n_S$), interaction parameter ($c^2$), and the epoch of
reionization measured by the optical depth of the Universe to CMB
photons ($\tau$). We do not consider gravitational waves, massive
neutrinos, or universes with curved spatial sections. Effectively,
our parameter space is 8-dimensional. A brute force analysis of
this parameter space would prove computationally very expensive.
Instead, we chose to explore the parameter space using a Monte
Carlo Markov chain (hereafter, MCMC) method, where the parameter
domain is more heavily explored close to the best fit values. This
method speeds up the calculation and at the same time explores
fully the likelihood function and allows to determine accurately
the model parameters and its error bars.  We implemented the
method as described in Ref. \cite{verde} which was first proposed
by Gelman and Rubin \cite{gelman-rubin}. For each data set we run
4 different MCMCs initiated at different regions of the parameter
space. We followed the same practice than the WMAP team: the chain
automatically stopped when a given degree of convergence ($R\le
1.2$ in the notation of \cite{verde}) was attained and when, at
least, $40,000$ models were computed.

\begin{center}
\begin{table}[t]
\begin{tabular}{|p{.5cm}p{5.0cm}p{3.0cm}p{3.0cm}p{3.0cm}|}
\hline &$40\le H_0\le 100$ km/s Mpc$^{-1}$&$0\le\Omega_ch^2\le
1$&$0\le\Omega_bh^2\le 1$&$-1\le w_x\le -0.5$
\vspace*{.2cm} \\
&$0.5\le n_s\le 1.5$ &$0.5\le A_s\le 1.5$&$0.1\le e^{-2\tau}\le
1.5$ & $-9\le\log c^2\le -1$
\vspace*{.2cm} \\
\hline
\end{tabular}
\caption[]{Priors on different parameters for WMAP, SDSS data
analysis. MCMCs are constrained to take values within those
intervals. The amplitude of the matter power spectrum $A_s=1$ is
the amplitude fixed by the COBE normalization.}
\label{tab:priors1}
\end{table}
\end{center}

\subsection{Constraints on IQMs from WMAP 3yr data}

We run 4 MCMCs comparing models with WMAP 3yr data, using the same
likelihood code than the WMAP team which is available at the
Lambda Archive \cite{lambda-archive}. Table \ref{tab:priors1}
shows the intervals where parameters are let to vary. These priors
are conservative in the sense that they include all known
measurements and we do not expect these limits to have any effect
on our final results. Still, we first made a rough exploration of
the parameter space to ensure that the maximum of the likelihood
function remains far away from the surface of the parameter space
volume, so the likelihood hypersurface is not affected by our
priors. We stop the program when the chains have satisfied
convergence and well-mixing criteria.

\begin{figure}[t]
\centering
\includegraphics[scale=0.8]{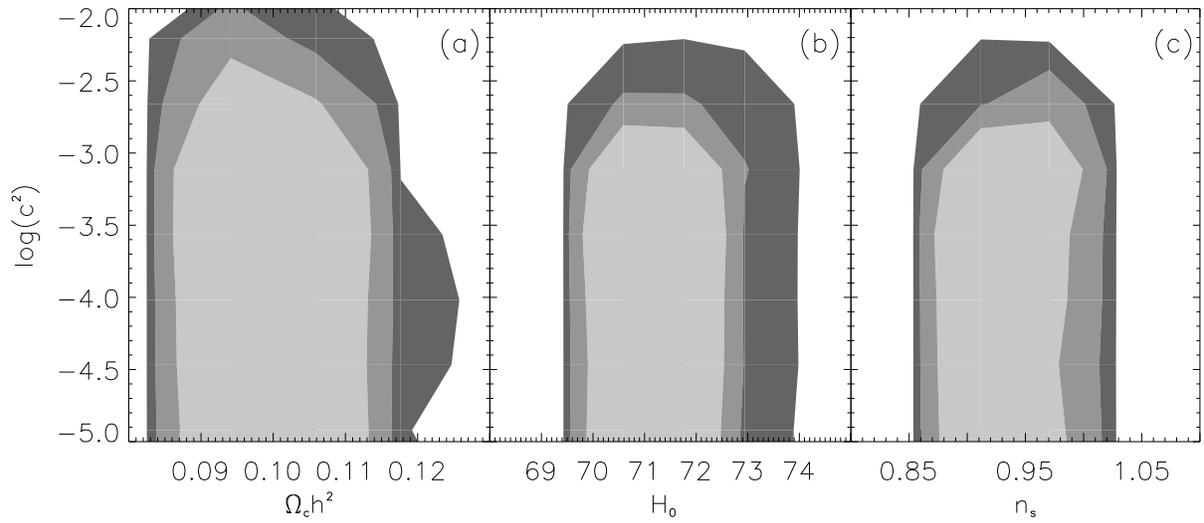}
\vspace*{-6cm}
\caption{ Joint confidence intervals at the 68\%,
95\% and 99.9\% level for pairs of parameters after marginalizing
over the rest. For convenience, the $c^2$ axis is represented
using a logarithmic scale and it has been cut at $c^2\le
10^{-5}$.} \label{fig:c2wmap3}
\end{figure}

\begin{figure}[t]
\centering
\includegraphics[scale=0.8]{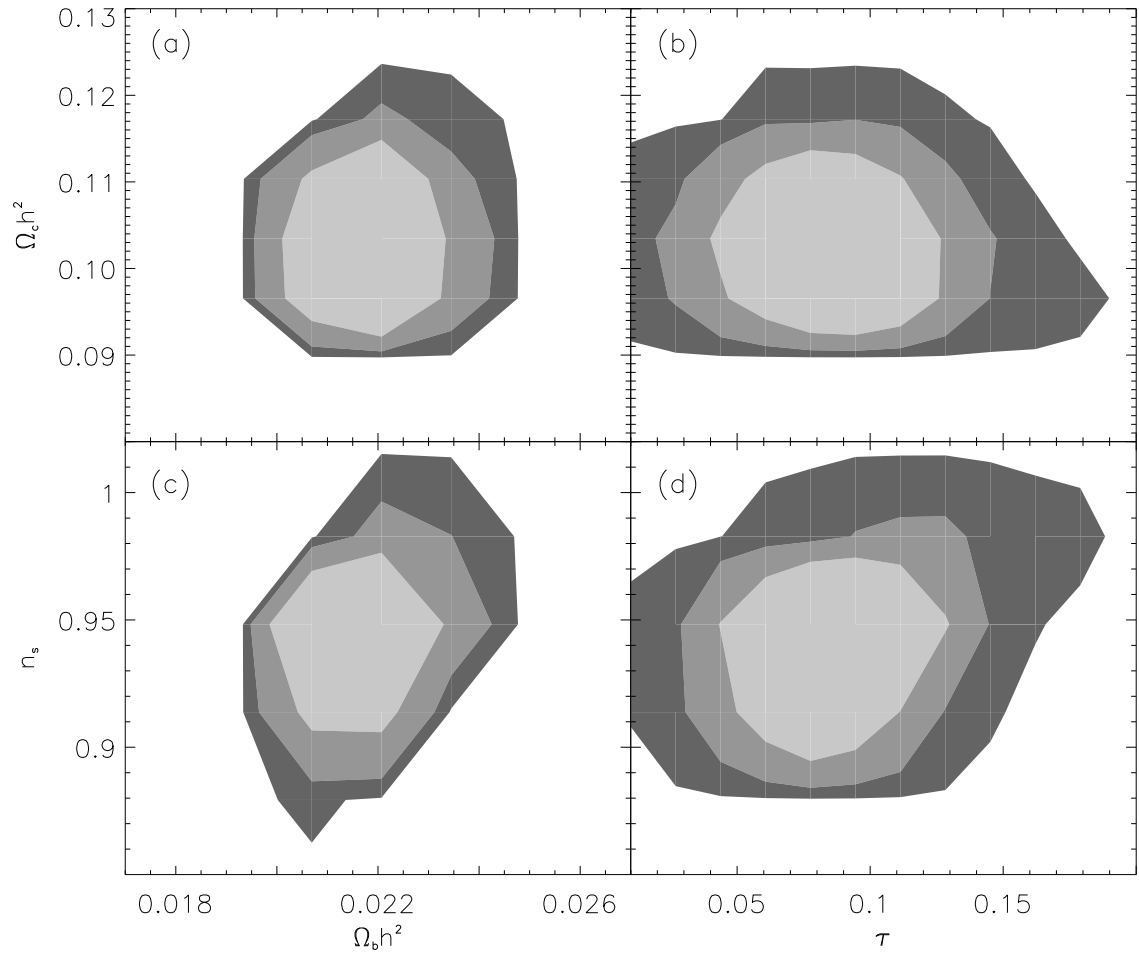}
\vspace*{0cm}
\caption{Joint confidence intervals at the same confidence levels as
in Fig. \ref{fig:c2wmap3} for pairs of parameters after marginalizing over the rest.}
\label{fig:wmap3}
\end{figure}

Figure \ref{fig:c2wmap3} presents the joint confidence intervals
at 68\%, 95\% and 99.9\% for pairs of parameters after
marginalizing over the rest. For convenience, the $c^2$ axis uses
a logarithmic scale and it has been cut at $c^2\le 10^{-5}$. In
the figure, contours run parallel to the vertical axis, meaning
that the data has only statistical power to set an upper limit on
the interaction parameter.  At the $1\sigma$ confidence level,
$c^2\le 2.3\times 10^{-3}$.  Figure \ref{fig:wmap3} shows the same
joint confidence contours for other pairs of parameters. In Table
\ref{exp-values} we presents the best fit values of the
cosmological parameters obtained from WMAP 3yr data alongside
their $1\sigma$ error bars. Those intervals are in rather good
agreement with WMAP 3yr results: the values on $\Omega_ch^2$ are
slightly more uncertain and on $H_0$ more precise. The small
differences are to be expected due to our different sampling of
the parameter space \cite{lewis}. The main difference lies in
$w_x$. For this parameter, we can only set upper limits.  The
reason is that we restricted our analysis to quintessence models,
$w_x\le -1$, and did not consider phantom models.

Let us remark that the shape of the likelihood function is not
altered by the introduction of a new degree of freedom. It is well
known that the CMB temperature anisotropies are highly degenerated
and different combination of parameters give rise to very similar
radiation power spectra. One would expect that one extra parameter
would introduce extra degeneracies and modify the overall shape of
the likelihood function. This is not the case  and it simply
reflects the fact that the best fit parameters are rather
insensitive to the interaction (see Fig. \ref{fig:c2wmap3}). An
obvious explanation is that WMAP data does not allow for large
departures from non-interacting models. The interaction changes
the depth of the potential  wells at the last scattering surface
but this effect cannot be counter-balanced by a suitable variation
of other cosmological parameters. The maximum of the likelihood
function is unaffected and the contours are not shifted.

\begin{table}
\begin{center}
\begin{tabular}{|c|c|c|}
\hline
\hspace*{.5cm}Parameter\hspace*{.5cm} &\hspace*{1cm} WMAP\hspace*{1cm}
&\hspace*{1cm} SDSS\hspace*{1cm} \\
\hline
$c^2$ & $\le 2.3\times 10^{-3}$  & $ \leq 4.8\times 10^{-3}$ \\
$\Omega_c h^2$ & $0.105^{+0.009}_{-0.012}$& $0.24^{+0.23}_{-0.13} $\\
$10^2\Omega_b h^2$ & $2.22\pm 0.07$ &$3.0^{+1.8}_{-2.7}$\\
$ w_x $ & $\le -0.90$ & $\le -0.6$\\
$H_0$ &   $71.3^{+1.2}_{-1.5}$ & $\cdots$\\
$n_s$ & $ 0.94^{+0.04}_{-0.05}$  & $0.85^{+0.15}_{-0.20}$\\
$\tau$& $0.09\pm 0.04$  & $\cdots$ \\
\hline
\end{tabular}
\caption{Mean values of cosmological parameters and their
$1\sigma$ errors obtained by fitting WMAP 3yr and SDSS LRGs data.
For $w_x$ only upper limits at the $1\sigma$ confidence level are
given.  Values not quoted are weakly or not constrained by the data.}
\label{exp-values}
\end{center}
\end{table}

\subsection{Constraints on IQMs from SDSS and 2dFRGS data}
The matter power spectrum has been recently derived from two
different galaxy catalogs: 2-degree Field Red Galaxy Survey
(2dFRGS) \cite{2df}, that comprises a sample of more than $2\times
10^5$ galaxies with measured redshifts, and the Luminous Red
Galaxy sample (LRGs) of the Sloan Digital Sky Survey (SDSS)
\cite{sdss}. A larger study, using the SDSS Data Release 5, and
comprising more than half a million galaxies showed similar
results \cite{sdss2}. In \cite{sdss}, the power spectrum from the
SDSS was computed producing uncorrelated minimum-variance
measurements in 20 $k$-bands of the clustering power and its
anisotropy due to redshift-distortions, with narrow and
well-behaved windows functions in the range $0.01\, h/\mbox{Mpc}$
$<k<\, 0.2\,h/\mbox{Mpc}$. We used these data to fit the
theoretical power spectrum, $P(k)$, calculated with a modified
CMBFAST software and thus constrain the parameters of the
cosmological model. Table \ref{tab:priors1} shows the priors
imposed on the parameters in our MCMCs runs. In addition, we
allowed the bias parameter to vary in the range $0.5\, < b <
\,2.5$. We run 4 different MCMCs till the criteria of convergence
and well mixing were met.

\begin{figure}[t]
\centering
\includegraphics[scale=0.8]{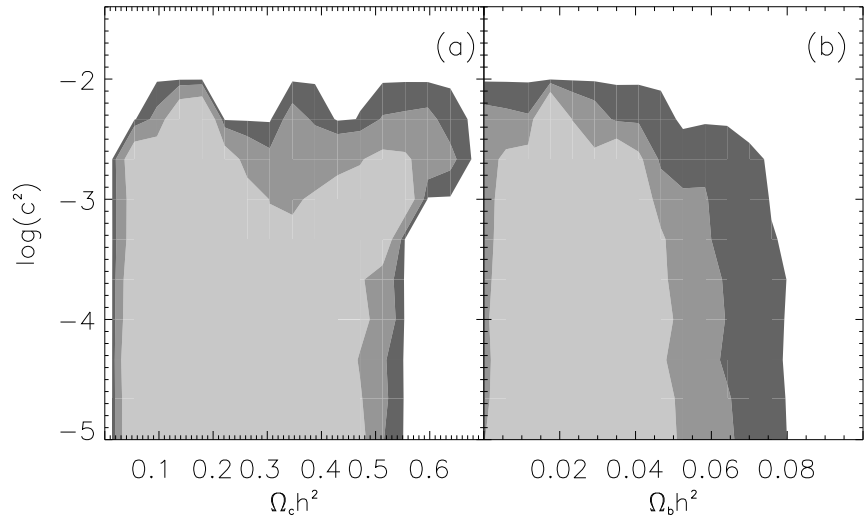}
\vspace*{-6cm}
\caption{Contours as in Fig. \ref{fig:c2wmap3}
obtained using the SDSS  matter power spectrum data.}
\label{fig:sdss_c2}
\end{figure}

Figure \ref{fig:sdss_c2} shows the joint confidence contours of
$c^2$ with other representative parameters. Like in Fig.
\ref{fig:c2wmap3}, contours run almost parallel to the vertical
axis. In Table \ref{exp-values} we give the cosmological
parameters measured from this data set. We derived an upper limit
of $c^2\leq 4.8\times 10^{-3}$ at $1\sigma$ confidence level. This
constraint is weaker than the one provided by WMAP 3rd year data,
but applies to a different range in redshift. This bound sets an
upper limit on the value of  $c^2$ averaged over the whole length
of the matter dominated epoch.  Varying $H_0$ or the total matter
content changes the matter radiation equality and shifts the
maximum of $P(k)$ to larger scales, but it is around the maximum
where the data have less statistical power. Matter and baryons
change the amplitude and location of baryon acoustic oscillations
but useful constraints can only be obtained by separating the
physics of the oscillations from that governing the overall shape
of the power spectrum \cite{percival-2007} loosing the information
on the interaction. As a result, we obtained no useful constraints
on cosmological parameters except on $c^2$. We also run a set of
MCMCs for the 2dFGRS data, but while the values of the main
cosmological parameters were in very good agreement with those of
Table \ref{exp-values}, the constraints were even poorer than
those obtained from the SDSS data, so we do not present them here.

\subsection{Constraints from SNIa data}
\begin{figure}[t]
\centering
\includegraphics[scale=.8]{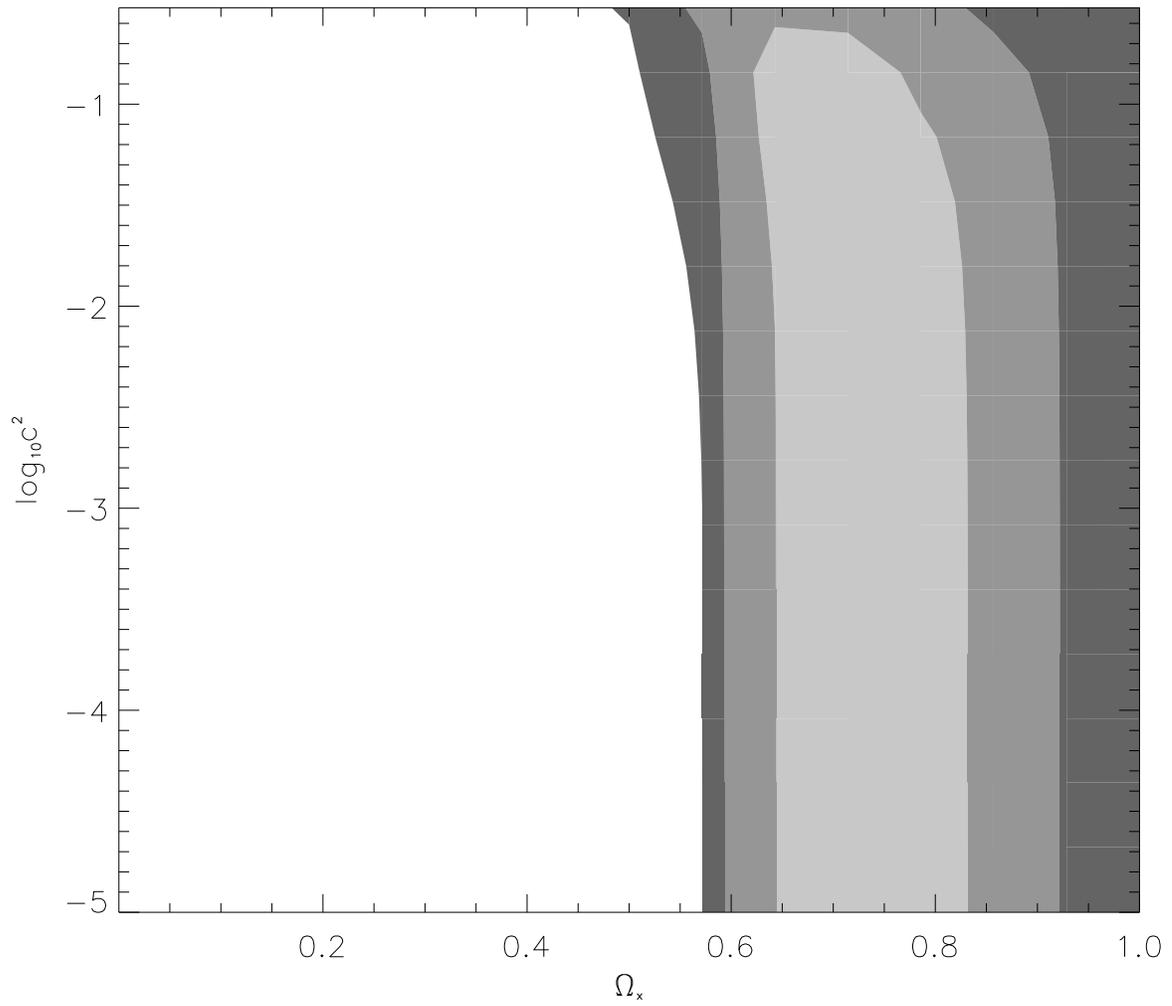}
\vspace*{0cm} \caption{Contours as in Fig. \ref{fig:c2wmap3}
obtained by fitting the luminosity distance IQM to the ``Gold"
sample of SNIa data of Riess {\em et al.} and SNLS data of Astier
{\em et al.} \cite{riess}.} \label{fig:likeSN}
\end{figure}

At low redshifts, only the dark matter and dark energy densities
contribute substantially to the dynamical evolution of the
Universe. Since the interaction alters their relative ratio, it
affects the actual distance to objects. Distance measurements,
like the luminosity distance, $d_{L}$, obtained from standard
candles like the SNIa, are sensitive to the dark energy
parameters, $\Omega_{x}$, $w_{x}$, and the strength of the
interaction, $c^2$. The likelihood function is given by
\cite{riess},
\\
\begin{equation}\label{chiSN}
{\cal L} \propto \exp(-\frac{1}{2}\chi^2), \qquad
\chi^2(H_0,w_x,\Omega_x,c^2)=\sum_i\df{\left(\mu_{the}(z_i;H_0,w_x,\Omega_x,c^2)-\mu_{obs,i}
\right)^2}{\sigma^{2}_{\mu_{obs,i}}+\sigma^{2}_{v}}\, .
\end{equation}
\\
In this expression, $\mu=5\ln{d_L}+25$ is the distance modulus,
$\sigma^{2}_v$ is the uncertainty on the supernova redshift (in
units of the distance modulus) due to its intrinsic motion and the
peculiar velocities of the parent galaxy and
$\sigma^{2}_{\mu_{obs,i}}$ is the error in the determination of
the distance moduli due to photometric uncertainties. Subindexes
$the$ and $obs$ indicate theoretical and observed uncertainties,
respectively. Since the intrinsic luminosity of SNIa is not known,
the distance moduli are insensitive to $H_0$, only 3 parameters
($\Omega_x$, $w_x$, and $c^2$) are needed to characterize a model.
We carried out a brute force analysis sampling homogeneously the
parameter space: ($-9\le\log c^2\le -1$, $0\le\Omega_x\le 1$,
$-1\le w_x\le -0.5$); each interval was subdivided in 10 steps.
Figure \ref{fig:likeSN} shows the confidence intervals for $c^2$
vs $\Omega_{x}$ after marginalizing over $w_{x}$. We remark that,
in this case, the data is almost insensitive to  $c^2$, as we
could have expected from Fig \ref{fig:cmb3rd}(c).

\subsection{Bayesian model selection}
To solve the coincidence problem a decay of  DE into DM was
postulated and, consequently, a new parameter, $c^{2}$, was
introduced to gauge the strength of the coupling. In this
subsection we explore whether the existence of this coupling
(i.e., a non-vanishing $c^{2}$ value) it is also justified by a
statistical description of the data. The introduction of further
parameters will lead to a better fit to the data. This must be
balanced against the loss of predictive power of the theory. The
Bayesian Information Criteria (defined as $BIC=\chi^2+k\log N$
\cite{liddle}, where $\chi^2=-2\log{\cal L}$, $k$ the number of
free parameters, and $N$ the number of data points), penalizes the
inclusion of additional parameters to describe data of small size.
For the WMAP 3yr data $\chi^2(c^2 = 0)= 11252.06$ and $\chi^2
({\rm best \, model}) = 11251.85$, so the interacting quintessence
model for $c^{2} = 2.3 \times 10^{-3}$ yields a slightly better
fit to data than the corresponding non-interacting model. However,
the application of the Bayesian criterion gives $BIC({\rm best\,
model})= 11269$ and $BIC(c^2=0) =11266$, so the introduction of
this parameter is not statistically encouraged. This results
disagree with our analysis of WMAP 1yr data \cite{olivares1}. It
was the unexpected consequence of our choice of priors: we fixed
all models to agree with the COBE/DMR normalization. Since WMAP
1yr concordance model preferred a lower normalization, it was
heavily penalized compared with interacting models. This
discrepancy shows the importance of priors in Bayesian analysis.
As argued in \cite{linder}, model selection is not a substitute
for parameter fitting; a lower BIC does not imply the data rejects
interacting models, just makes them unlikely. However, they still
have a theoretical advantage since they can help explain the
coincidence problem, while non-interacting models cannot.

\section{Conclusions}
Let us briefly point out that while the IQM model does not solve
the coincidence problem in full (as it cannot predict the present
ratio of dark matter to dark energy \footnote{To best of our
knowledge, no model predicts this value.}) it substantially
alleviates it. Indeed, if we take initial conditions at the Planck
scale, in a model without interaction the ratio $r$ at that epoch
is about $10^{96}$. If $c^2 \sim 10^{-3}$, a value compatible with
WMAP upper limit then, asymptotically, $r(t\gg t_0)\sim 10^3$,
where $t_0$ is the present time.  Initially $r(t\ll t_0)\sim
10^{-3}$, so, while in a $\Lambda$CDM model  the initial
conditions have to be fine-tuned by $96$ magnitude orders to get
the present accelerated expansion, in  the IQM the difference in
orders of magnitude is just $6$. A further advantage is that the
model is not limited by the results of Ref.
\cite{amendola-challenges} since it successfully predicts that the
Universe undergoes three successive cosmic eras of expansion,
namely, radiation, matter and dark energy dominance. If the
interaction parameter were  $c^2\sim {\cal O}(10^{-1})$ or larger,
it could undergo an era of baryon dominance, but this is not a
stable solution and, in any case, $c^{2}$ values larger than $2.3
\times 10^{-3}$ do not appear compatible with CMB data.

\begin{figure}[t]
\centering
\includegraphics[scale=0.8]{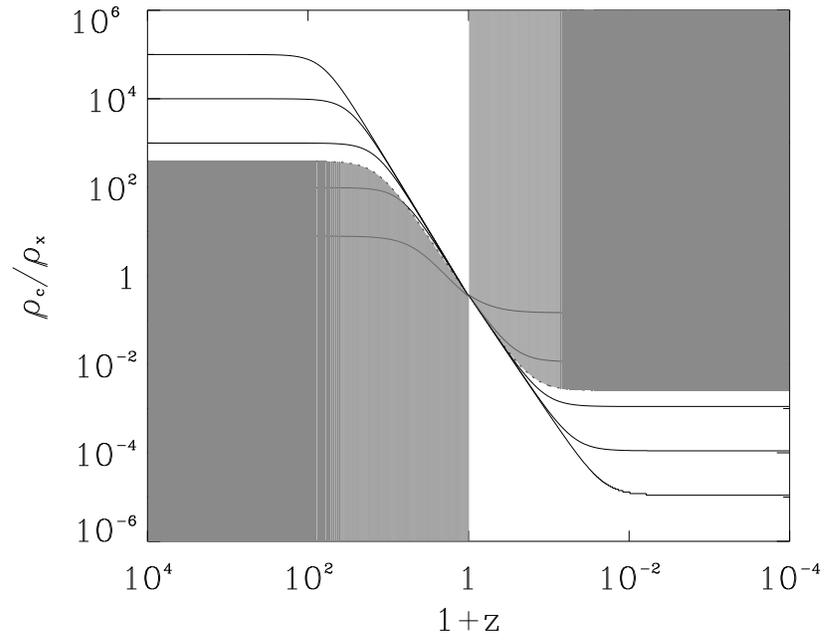}
\vspace*{0cm}
\caption{Ratio DM/DE vs redshift.  The $c^{2}$ values on the
curves from top to bottom (on the left hand side) are $10^{-5}, \,
10^{-4},\, 10^{-3}$, and $2.3 \times 10^{-3}$. The shaded areas
indicate the two regions  excluded by the WMAP 3yr upper limit on
$c^2$. } \label{fig:excluded}
\end{figure}

The existence of a late accelerated expansion of the Universe
requires $w_x < -1/3$, as in noninteracting models. Whether or not
the present value of the ratio $r$ coincides with the final
attractor value depends on the parameters $c^2$ and $w_x$. These
are unknowns that can be constrained by observation. Our analysis
shows that the CMB, matter power spectrum and luminosity distance
data only set upper limits on the DM/DE coupling and,
statistically, the introduction of this extra parameter is not
favored by the data. At present, CMB data sets the tightest upper
limit on the strength of the interaction ($c^{2} \leq 2.3 \times
10^{-3}$). As Fig. \ref{fig:excluded} reveals, this upper bound
can be translated into an excluded region (shaded area) for the
dark matter to dark energy ratio, $r$. Solid lines, from top to
bottom, correspond to $c^2=10^{-5},10^{-4},10^{-3}$ and $2.3\times
10^{-3}$ that bounds the shaded area. As the figure shows, the
smaller $c^2$, the longer it takes to reach the attractor
solution.  With WMAP 3yr constraint, the attractor solution will
be reached only in $10^{11}$ yr from now.  Our upper limits on
$c^2$ put pressure on this type of interacting models that,
combined with the more generic results of
\cite{amendola-challenges}, show the difficulties that interacting
models encounter into explaining the coincidence problem. If the
new WMAP 5yr data release sets an even stronger constraint on
$c^2$, the IQM will loose much of its appeal.

\acknowledgments { This research was partly supported by the
Spanish ``Ministerio de Educaci\'on y Ciencia" under Grants
FIS2006-12296-C02-01, BFM2000-1322 and PR2005-0359, the ``Junta de
Castilla y Le\'{o}n" (Project SA010C05) and the ``Direcci\'{o}
General de Recerca de Catalunya" under Grant 2005 SGR 00087.}

\newpage

\newpage

\newpage

\newpage

\newpage

\newpage
\newpage

\end{document}